\documentclass[journal]{IEEEtran} 
\usepackage{citesort}
\usepackage{amsmath,amsthm}
\usepackage{algorithmic,algorithm}
\usepackage{array}
\usepackage{amsfonts}
\usepackage{graphicx,color,overpic,psfrag,epsfig}
\usepackage{bm}
\usepackage{footmisc}
\usepackage{xcolor}
\usepackage{color}
\usepackage{subfigure}
\usepackage{epstopdf}
\usepackage{amssymb}

\newcommand{\qW}{{\bf W}}

\newcommand{\qq}{{\bf q}}

\begin{document}

\title{Model-Driven Beamforming Neural Networks} 

\author{\IEEEauthorblockN{Wenchao Xia,  Gan Zheng,  Kai-Kit Wong, and Hongbo Zhu}
\thanks{Wenchao Xia and Hongbo Zhu are with the Jiangsu Key Laboratory of Wireless Communications, and also with the  Engineering Research Center of Health  Service System Based on Ubiquitous  Wireless  Networks,   Ministry of Education, Nanjing  University  of Posts and Telecommunications, Nanjing 210003, China, e-mail:  $\{\rm 2015010203, hbz\}@njupt.edu.cn$}
\thanks{Gan Zheng is with the Wolfson School of Mechanical Electrical and Manufacturing  Engineering,  Loughborough  University,  Leicestershire  LE11  3TU, U.K., e-mail:  $\rm g.zheng$@$\rm lboro.ac.uk$}
\thanks{Kai-Kit Wong is with the Department of Electronic and Electrical Engineering, University College London, London WC1E 7JE, U.K., e-mail: $\rm kai$-$\rm kit.wong@ucl.ac.uk$}
\thanks{\copyright 20XX IEEE.  Personal use of this material is permitted.  Permission from IEEE must be obtained for all other uses, in any current or future media, including reprinting/republishing this material for advertising or promotional purposes, creating new collective works, for resale or redistribution to servers or lists, or reuse of any copyrighted component of this work in other works.}}

\maketitle
\begin{abstract}
Beamforming is evidently a core technology in recent generations of mobile communication networks. Nevertheless, an iterative process is typically required to optimize the parameters, making it ill-placed for real-time implementation due to high complexity and computational delay. Heuristic solutions such as zero-forcing (ZF) are simpler but at the expense of performance loss. Alternatively, deep learning (DL) is well understood to be a generalizing technique that can deliver promising results for a wide range of applications at much lower complexity if it is sufficiently trained. As a consequence, DL may present itself as an attractive solution to beamforming. To exploit DL, this article introduces general data- and model-driven beamforming neural networks (BNNs), presents various possible learning strategies, and also discusses complexity reduction for the DL-based BNNs. We also offer enhancement methods such as training-set augmentation and transfer learning in order to improve the generality of BNNs, accompanied by computer simulation  results and testbed results showing the performance of such BNN solutions.
\end{abstract}

\section*{Introduction}
The ever growing demand for mobile data as a result of new lifestyles and innovative applications has continuously pushed the limits of today's mobile communication networks and created new challenges. 5G is the most recent and largest collective effort to bring the technology up to speed for the next 10 years or so. One core technology that has appeared in recent generations of mobile communication technology and will continue to have its presence in future generations is the multiuser multiple-input multiple-output (MIMO) system. This technology provides extraordinary spectral and energy efficiency by using the spatial degree of freedom that can be scaled up by having more antennas.

Beamforming in multiuser MIMO is a popular and excellent method for dealing with interference, especially in the downlink. There has been a rich body of literature on that, spanning from sum rate maximization \cite{shi2011an}, to signal-to-interference-plus-noise ratio (SINR) balancing  \cite{schubert2004solution}, to quality-of-service (QoS) constrained base station (BS) transmit power minimization \cite{yu2007transmitter} and among others. The algorithms to find the optimal and even suboptimal solutions to these beamforming problems usually require iterative optimization procedures. The complexity and resulting latency make those techniques problematic for real-time applications because wireless channel fading changes rapidly in the order of milliseconds and expensive iterative procedures will render the obtained solutions invalid if the channel state information (CSI) becomes obsolete. Heuristic non-iterative solutions such as zero-forcing (ZF) and regularized ZF (RZF) beamforming exist but come at the price of performance loss.

Recent developments in deep learning (DL) have given this problem a new hope. DL is well known to be a generalizing technique that can produce an effective solution to a complex problem at relatively low complexity if it is sufficiently trained. The approach means that most complexity is shifted to offline training an artificial neural network (NN) with a large dataset \cite{chang2018learn}. An online solution can then be obtained by going through a trained NN generalizable from the dataset, with some simple linear and standard nonlinear operations  \cite{xia2019deep}. Researchers have applied DL to network deployment and planning, resource management,  and network operation and maintenance. Specific to the physical layer, DL applications include modulation recognition, channel estimation and detection, and encoding and decoding. These applications demonstrate the potential of applying DL in wireless communication networks.

Nevertheless, using DL for beamforming is not straightforward for many reasons. First of all, the number of variables for beamforming depends on the number of users and the number of antenna elements which are usually large. The beamforming problem's high-dimensionality will cause issues in prediction complexity and errors. A common method in the existing works, such as \cite{long2018data}, is codebook-based beam selection, but tends to suffer certain performance loss. Further, many beamforming optimization problems are non-convex, which makes it difficult to have high-quality training examples, if they are to be solved using a supervised learning approach. It is also unclear if unsupervised learning can be useful for beamforming. In addition, artificial NNs are often oversized, meaning that a lot of neurons are redundant, resulting in unnecessarily high complexity and memory cost. Also, the generality of DL-based solutions can be very limited and a new NN will need to be trained if the parameters of the wireless communication networks change. Understanding the limitations of {\em data-driven} DL for beamforming, this article introduces the {\em model-driven} beamforming NNs (BNNs) as a new means to utilize DL for beamforming optimization.

The remainder of this article is organized as follows. We will begin by reviewing and comparing the data-driven and model-driven BNNs based on the convolutional NN (CNN) structure. Then we will focus on the model-driven BNN framework in which a signal processing (SP) module is brought into the NN module to enhance the process of feature extraction. Rather than solely predicting the beamforming solution, the proposed BNN framework allows to predict the key features according to expert knowledge with much reduced dimension. We will use the SINR balancing problem as  an example to illustrate the operation of this approach. Afterwards, several enhancement techniques, such as a hybrid learning strategy that combines unsupervised and supervised learning to deal with the issues for the lack of labelled examples, and NN trimming and compression that reduce the complexity and memory cost of the NN module, are proposed. Then we will discuss the use of training-set augmentation and transfer learning to improve the generality of DL-based BNNs before presenting simulation and testbed results and concluding the article.

\section*{BNNs}
There are two main types of DL-based BNNs: data-driven and model-driven ones. The main difference is that the former takes the NN as a black box, while the latter introduces  a specific SP module  into the NN \cite{he2018model}.

\begin{figure*}
\includegraphics[width=0.75\textwidth]{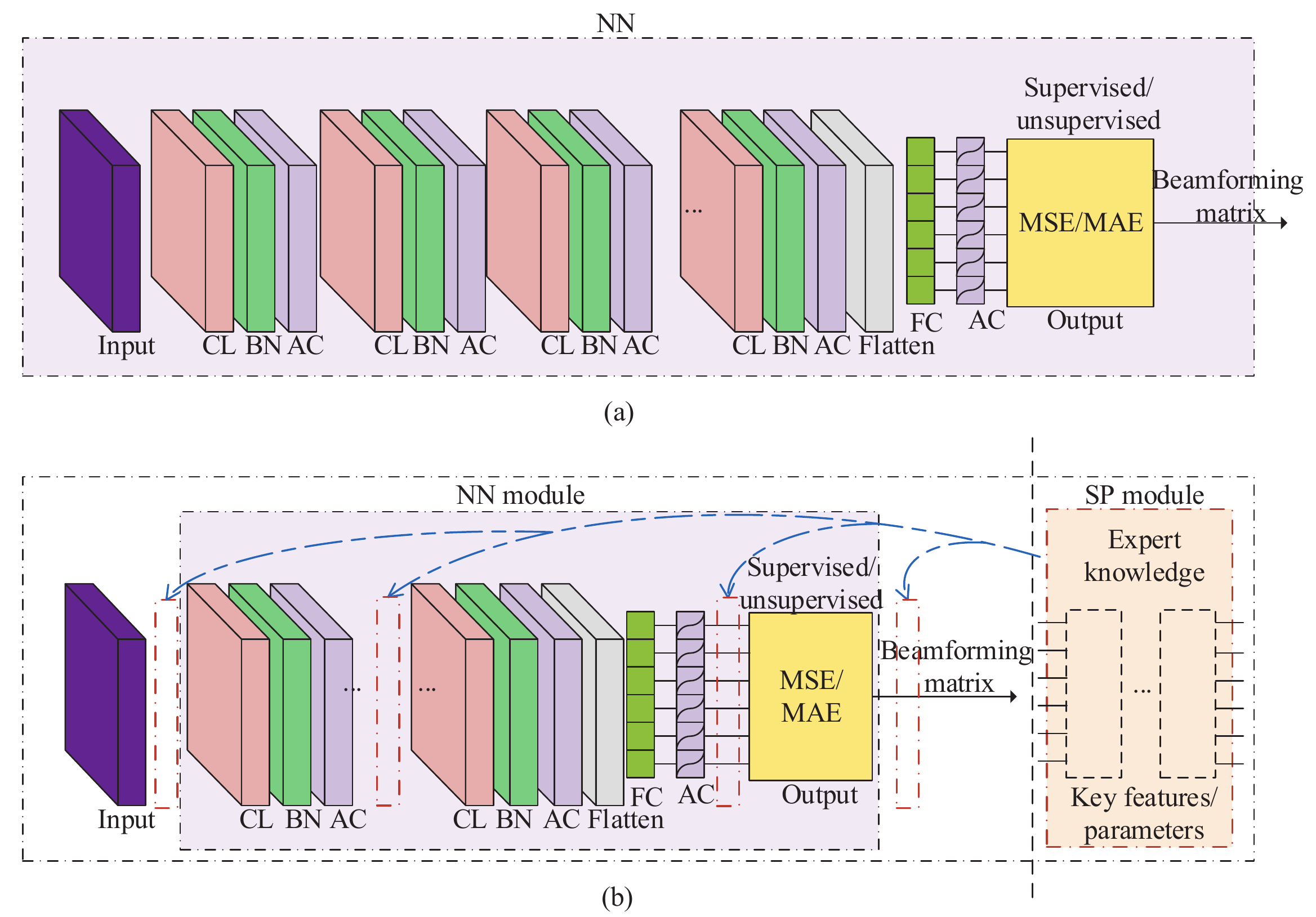}
\centering
\caption{Examples of a) data-driven and b) model-driven BNNs. The NN module is composed of an input layer, several convolutional (CL) layers, batch normalization (BN) layers, activation (AC) layers, a flatten layer, a fully-connected (FC) layer, and an output layer, whereas the key features and the functional layers in the SP module are specified by expert knowledge.}\label{data_driven_model_driven}
\end{figure*}

\subsection*{Data-Driven Architecture}
As shown in Fig.~\ref{data_driven_model_driven}a, a data-driven BNN follows the  structure of a CNN with an input layer, an output layer, and some hidden layers. The input layer takes real-valued channel coefficients as inputs to perform convolutional operations with kernels in the convolutional layers for feature extraction. The activation layers serve to introduce non-linearity into the NN, allowing it to capture complex functional mappings, and are also useful in mitigating the vanishing gradient problem for training the NN. The batch normalization layers are there to reduce the probability of over-fitting and enable a higher learning rate to accelerate convergence.

In the data-driven approach, the BNN acts like a black box and the functional establishment of the BNN relies heavily on both the quality and quantity of training samples. Moreover, the data-driven BNN is blind to any specialized signal structures, does not have the same  computational efficiency and the performance is often inferior to that of traditional SP methods. This is because traditional SP methods are crafted according to the prior expert knowledge, such as solution structures, uplink-downlink duality, models, and the properties of signals. Such a priori expert knowledge acquired from extensive research in the literature over the past decades, is expected to be highly useful and should be utilized \cite{zappone2019model}.

Crafting a complete SP solution using expert knowledge is however extremely difficult and sometimes impossible. It thus makes sense to combine the SP methods with the NN approach to reap the benefits of both sides \cite{zhang2018hybrid}. This then gives rise to the model-driven BNNs, which we will discuss next.

\subsection*{Model-Driven Architecture}
\begin{figure*}
\includegraphics[width=0.75\textwidth]{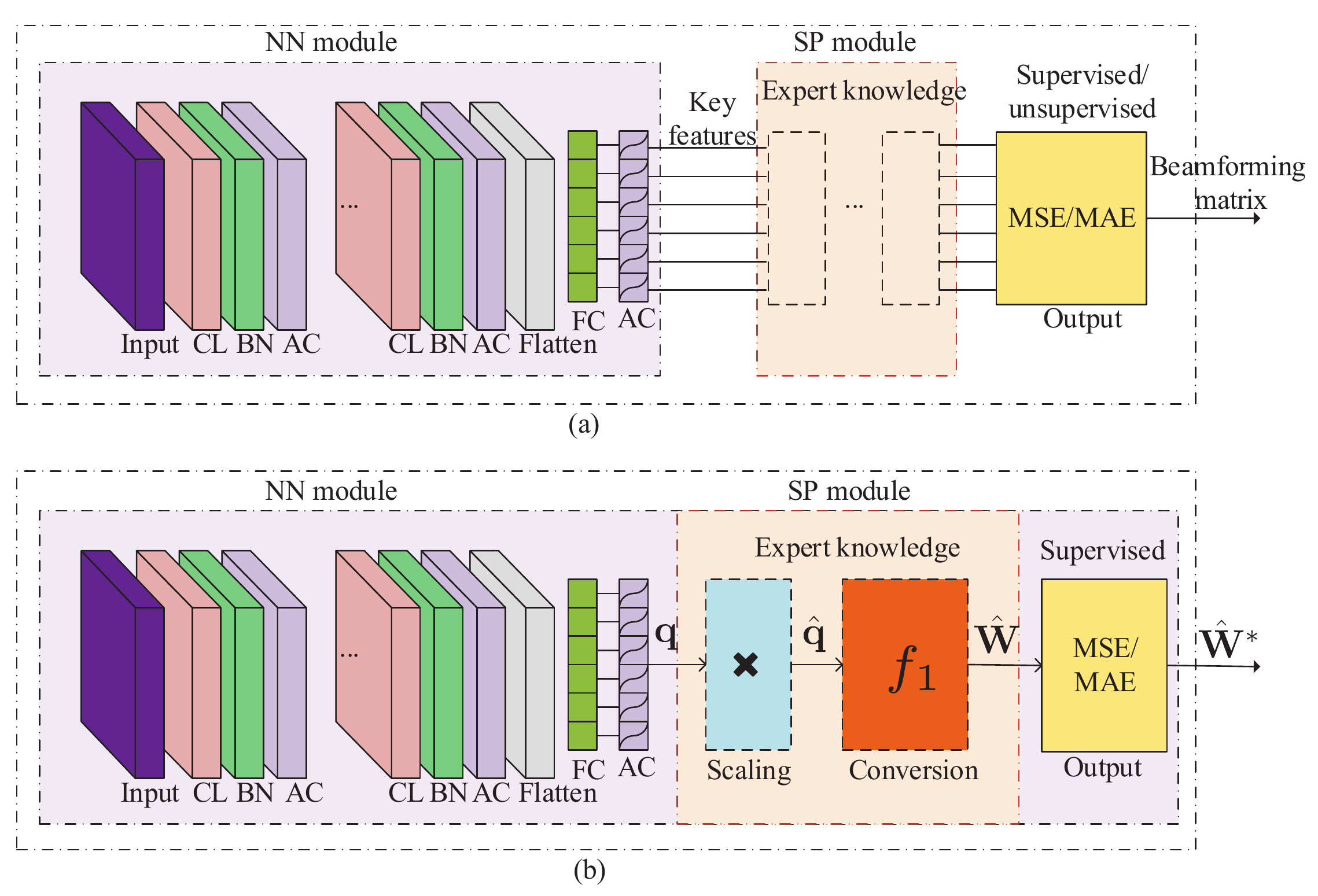}
\centering
\caption{a) A BNN framework and b) a BNN example for problem \textbf{P1}. CL: convolutional, BN: batch normalization, AC: activation, and FC: fully-connected.}
\label{framwork_example}
\end{figure*}

Different from the data-driven version, the proposed model-driven BNN has a specific SP module to utilize prior expert knowledge, as illustrated in Fig.~\ref{data_driven_model_driven}(b). The SP module can be positioned either before or after the NN module as a pre-processing or post-processing block. Inside the SP module are the functional layers that are designed according to prior expert knowledge of beamforming  problems, which is problem-specific and has no unified form \cite{xia2019deep}. It is also possible to replace one or more layers in the ordinary NN module by the SP module to achieve better feature extraction ability \cite{zhang2018hybrid}. The parameters in the SP module can also be tuned in the training phase.

The purpose of the SP module is to map/convert key features designated by the expert knowledge to the target beamforming matrix before updating the NN module. Let us use an example to explain the design process of the SP module, assuming, for convenience, a model-driven BNN framework where the SP module is inserted into the NN module and placed before the output layer, as shown in Fig.~\ref{framwork_example}a.

The example considers the use of a BNN for a multiple-input-single-output (MISO) downlink, where the aim is to balance the SINR under a total power constraint, i.e.,
\begin{equation}
\textbf{P1:}\ \max_{\qW}\min_{1\leq k\leq K} \gamma_k,~\text{s.t.}~\|\qW\|^2_F\leq P_{\max},
\end{equation}
where $\qW\in \mathbb{C}^{N\times K}$ denotes the beamforming matrix, $K$ is the number of single-antenna users, $N$ is the number of BS antennas, $\gamma_k$ is the SINR of user $k$, $P_{\max}$ is the total power budget, and $\|\cdot\|_F$ denotes the Frobenius norm.

The optimal solution to problem \textbf{P1} can be obtained by the algorithm in \cite[Table 1]{schubert2004solution}, using an iterative process that comes with high complexity and delay. Predicting the beamforming matrix $\qW$ directly using a data-driven BNN approach on the other hand will lead to a high prediction error since the number of elements in $\qW$ depends on both  the number of users and the number of BS antennas. To circumvent this, expert knowledge regarding uplink-downlink duality can be specified through the functional layers of the SP module of the model-driven BNN.

According to the duality theory in \cite[Theorems 1 and 3]{schubert2004solution}, both the uplink and downlink have the same achievable SINR region under the given total power constraint and the target SINRs are also achieved by the same set of normalized beamforming vectors. Thus, instead of solving problem \textbf{P1} directly, most existing works rightfully resort to its equivalent uplink problem which is easier to handle. In particular, suppose that we have for the equivalent uplink problem the optimal power allocation vector $\qq^{\ast}$ and the optimal normalized beamforming matrix $\tilde{\qW}^{\ast}$ with the same power budget $P_{\max}$. $\tilde{\qW}^{\ast}$ is also a function of $\qq^{\ast}$. It is then known that the optimal downlink beamforming matrix $\qW^{\ast}$ is a function of $\tilde{\qW}^{\ast}$. As a result, the optimal solution to problem \textbf{P1} is a function of $\qq^{\ast}$, i.e., $\qW^{\ast}=f_1(\qq^{\ast})$, where $f_1(\cdot)$ maps $\qq$ to $\qW$ based on the results in \cite{schubert2004solution}. Such expert knowledge suggests that instead of predicting the high dimensional $\qW$ directly, we can predict the uplink power vector $\qq$ with much less variables.

Based on the BNN framework in Fig.~\ref{framwork_example}a, the model-driven BNN for the SINR balancing problem is shown in Fig.~\ref{framwork_example}(b), where the SP module is fulfilled with two functional layers: the scaling layer and the conversion layer. The scaling layer is used to ensure that the beamforming matrix meets the power constraint by multiplying the estimated $\hat{\qq}$ by a scaling factor, whereas the conversion layer is used to execute the function $f_1(\cdot)$. After recovering from the uplink power vector via the function $f_1(\cdot)$, the resulting beamforming matrix is then used to calculate the loss function and update the parameters of the NN training module until convergence.

\section*{Supervised vs. Unsupervised Learning}

\begin{figure}
\includegraphics[width=0.45\textwidth]{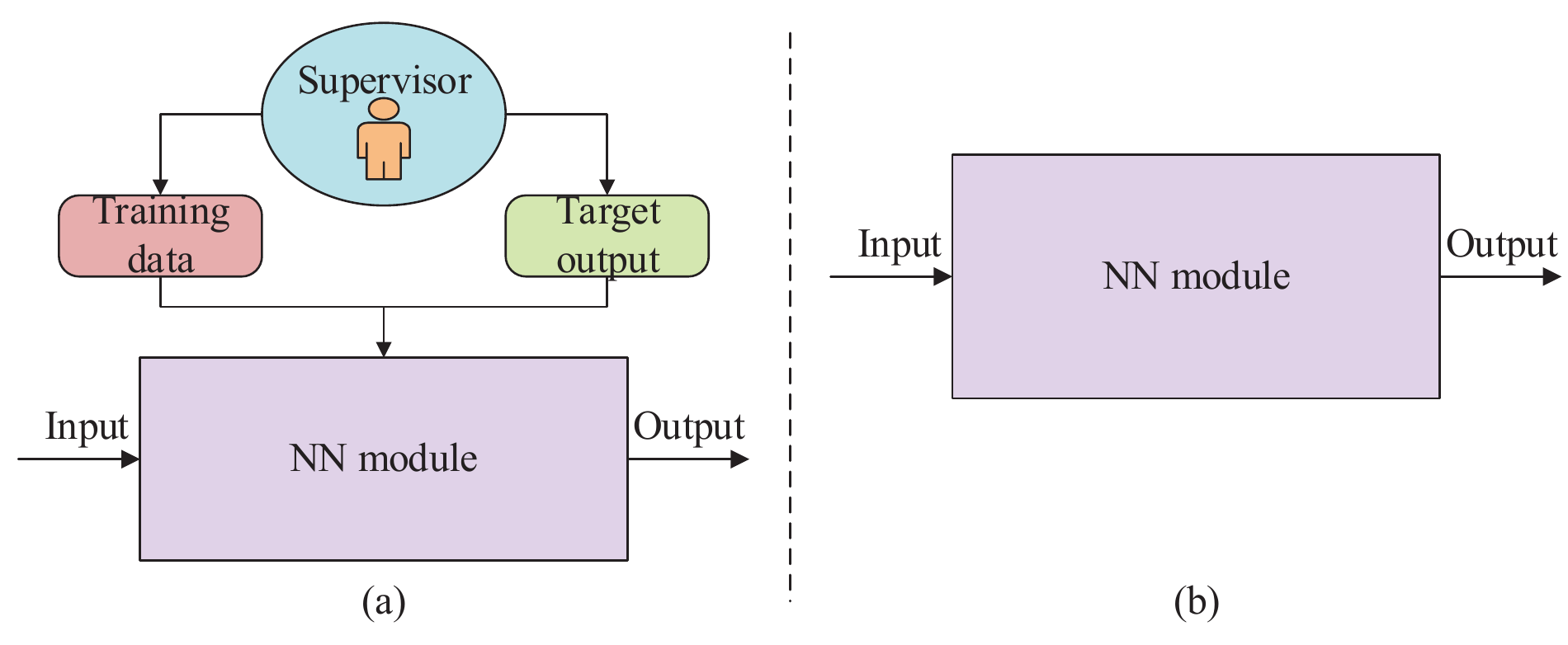}
\centering
\caption{a) Supervised learning and b) unsupervised learning.}\label{supervised_unsupervised_learning}
\end{figure}

Supervised learning and unsupervised learning are two very different approaches to training NNs. Supervised learning is based on a ground truth generalizable from labelled training samples, while unsupervised learning finds natural patterns from unlabelled data. In other words, the objective of supervised learning is to learn a mapping function that can well approximate the relationship between the input and desired output in the training samples. Unsupervised learning on the other hand infers the potential structure in the training data.


\subsection*{Supervised Learning for BNN}
Supervised learning generalizes the mapping between input and output based on the training samples using an NN representation. More samples improve the mapping accuracy. As both the input and the target output are known, the learning process is interpreted as being done by a ``supervisor'', see Fig.~\ref{supervised_unsupervised_learning}a. The NN module repeatedly makes predictions while the ``supervisor'' corrects the predictions based on the expected output in an iterative manner. Such a learning-correction process terminates until satisfactory performance is achieved.

For classification-type of beamforming problems, e.g., beam selection based on a codebook \cite{long2018data}, cross entropy loss is a well-known metric to quantify the prediction error. For regression problems of beamforming optimization, e.g., problem \textbf{P1}, on the other hand, mean squared error (MSE) and mean absolute error (MAE) are two common loss functions, where the former is the  average of squared distances and the latter is the average of absolute differences between the target and predicted outputs.

Due to the reliance on a large number of training samples, supervised learning is more suitable for cases where training samples are relatively easy to collect. For instance, for the SINR balancing problem under a total power constraint (i.e., problem \textbf{P1}) and the power minimization problems with QoS constraints, there exist computationally efficient algorithms in \cite{schubert2004solution} and \cite{yu2007transmitter} that can produce the optimal solutions as training samples, respectively. However, for some difficult beamforming problems, e.g., the sum-rate maximization problem under a total power constraint and SINR balancing problem  under per-antenna power constraints, the acquisition of training samples with optimal solutions  may be not easy or be at a much higher cost. Unsupervised learning provides an alternative for such beamforming optimization problems with limited samples.

\subsection*{Unsupervised Learning for BNN}
Unsupervised learning aims to infer the underlying structure of data without any label, and hence it is particularly suitable for exploratory analysis. For some beamforming optimization problems, no known algorithms can find optimal solutions, and the target output is unknown. In this case, MSE or MAE will not be suitable to measure the loss for updating the parameters of the NN module. Instead, the problem's original objective function may be used to construct the loss function. However, unsupervised learning with random initialization of parameters of BNNs usually suffers from a low convergence rate.

\subsection*{Hybrid Learning for BNN}
Supervised and unsupervised learning can complement each other, when used together. Such hybrid learning is a promising technique to achieve both great performance and accelerate the convergence rate of unsupervised learning. In the first stage, supervised learning is used for pre-training and then unsupervised learning will be used for further improvement in the second stage \cite{lee2018deep}. Getting the best of both learning methods, hybrid learning is an attractive approach known to achieve performance that is better than most existing heuristics.

Take the sum-rate maximization problem with a total power constraint as an example. Though no known efficient algorithm can obtain the optimal solutions, we can adopt the weighted minimum MSE (WMMSE) algorithm in \cite{shi2011an} to generate training samples with locally optimal solutions, which can then be used for supervised learning and pre-training. After that, the learned NN parameters are reserved for unsupervised learning and the loss function can be replaced by the reciprocal function of the sum-rate. Thus, the convergence rate of unsupervised learning is accelerated and hybrid learning can achieve at least the same performance as the WMMSE algorithm \cite{xia2019deep}.

\section*{Complexity Consideration}
Although BNNs are powerful, their computational complexity and memory cost can make them less attractive for resource-constrained hardware and equipment. Large-scale BNNs composed of massive neurons also have considerable energy consumption due to massive memory access and abundant computation. As a consequence, reducing the complexity is an important direction if the application of BNNs is to be practical. We next discuss ideas to reduce the complexity of DL-based BNNs in two aspects: the beamforming optimization problem itself and the NN module.

\subsection*{Complexity of the Optimization Problem}

\begin{figure*}
\includegraphics[width=0.75\textwidth]{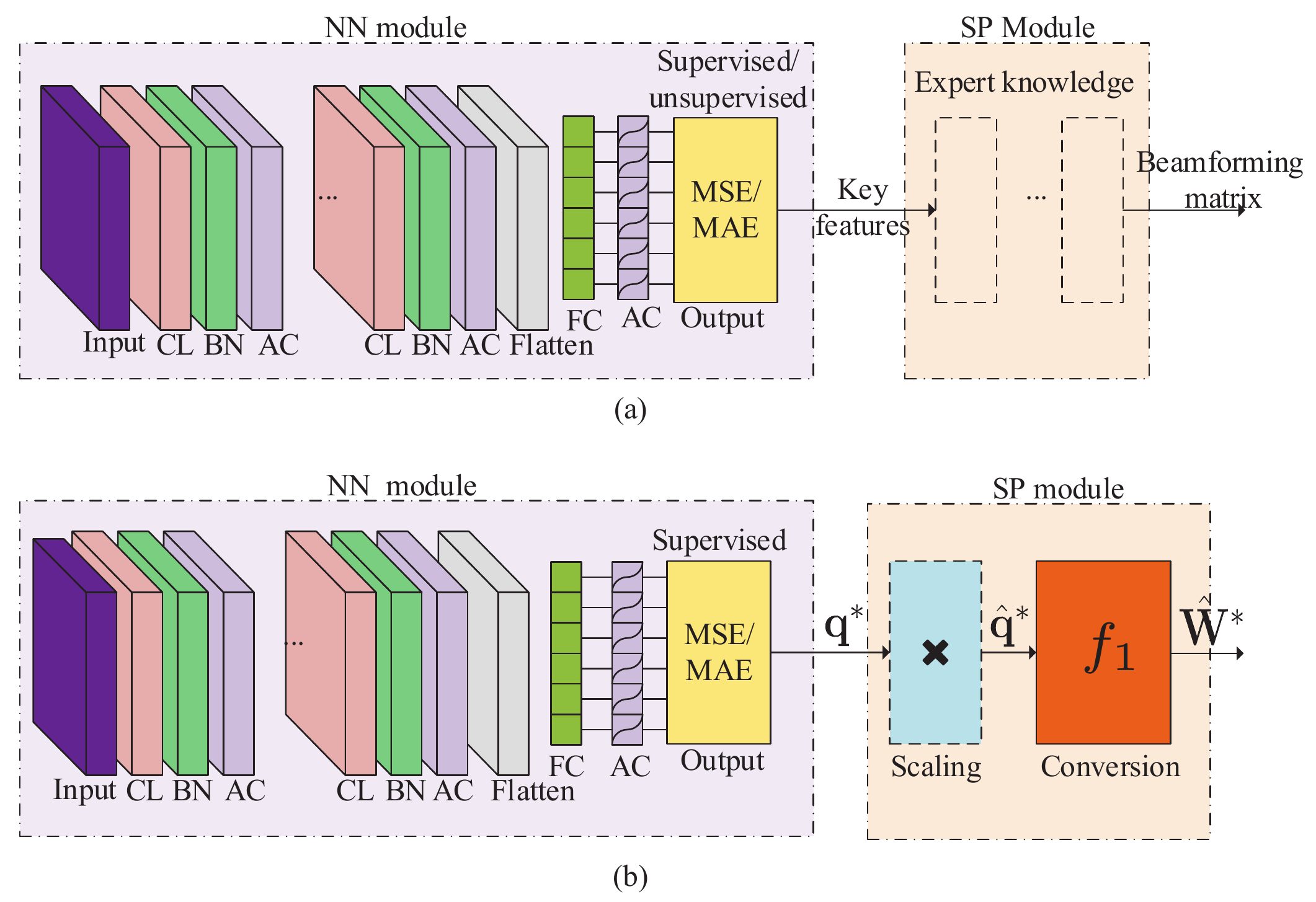}
\centering
\caption{a) The BNN framework with reduced prediction complexity and b) the corresponding BNN example for problems \textbf{P1} \cite{xia2019deep}. CL: convolutional, BN: batch normalization, AC: activation, and FC: fully-connected.}
\label{reduce_complexity}
\end{figure*}

The input and output of the NN are specified according to the beamforming optimization problem. One way to reduce the prediction complexity is to shrink the dimensions of the input and output of the NN. A straightforward strategy which takes the wireless channel coefficients as input and the beamforming matrix as output may give rise to high prediction complexity because of their dependence on the numbers of users and BS antennas. To remove the redundant information carried by the input of the NN module,  a promising scheme for beam selection problems is to take the locations of users and the BS as input instead of the channel coefficients. Another promising technique, which takes place at the output, is to predict the key features, but not the beamforming matrix, according to expert knowledge, which we elaborate in more details below.

It is possible to first predict some key features designated by prior expert knowledge and then recover the beamforming matrix from the predicted key features, as shown in Fig.~\ref{reduce_complexity}a. Different from the BNN framework in Fig.~\ref{framwork_example}a which predicts the beamforming matrix and updates the parameters of the NN module for minimizing the loss function of the predicted beamforming matrix, the BNN framework in Fig.~\ref{reduce_complexity}a proposed by our previous work \cite{xia2019deep} only predicts the key features and calculates the loss function based on the predicted key features until convergence. The final step, which is executed only once, is to use the functional layers in the SP module to retrieve the beamforming matrix. In contrast, those SP layers in Fig.~\ref{framwork_example}a are executed repeatedly during the entire training process. In addition, although the key features are abstract and problem-specific, the most important advantage is that the number of key features is often much less than the number of variables in the beamforming matrix.

Based on the BNN framework in Fig.~\ref{reduce_complexity}a, the solution to the corresponding BNN problem \textbf{P1} can be found with reduced complexity using the approach shown in Fig.~\ref{reduce_complexity}b \cite{xia2019deep}.

\subsection*{Complexity of the NN module}

There is no formal way to obtain the best number of layers and the best number of neurons for each layer in the NN module. In many cases, numbers such as 64, 128, and 512 are typically used but such empirically designed NNs are usually oversized \cite{hu2016network}.  In other words, many neurons may have very low activation regardless of the input and these weak neurons can be removed with little performance loss. Redundant neurons, if not removed, increase computational complexity, memory cost, and the probability of over-fitting, and are thus highly undesirable in terms of the NN performance.

To reduce the complexity of the NN module, we can first use trimming to prune all those connections with weights below a certain threshold and those with zero-activation neurons. Then we reduce the number of bits used to represent each weight via a compression technique and enforce weight sharing among different connections to reduce the number of weights. Finally, Huffman coding can be adopted to represent more common weights using symbols with fewer bits \cite{hu2016network}.

\section*{Generality Improvement}
In most existing works, the DL-based approaches to beamforming prediction can achieve very good performance but the NNs were trained with fixed wireless network parameters, meaning that the numbers of users and antennas, for example, are fixed. However, wireless networks are dynamic in nature. For example, user mobility means that users join and leave the network over time. Also, network operators may turn off/on a subset of BS antennas according to traffic load, leading to the variation of the number of serving antennas. The pre-trained model may  suffer from serious performance degradation and even become unusable because the dimensions of input and output do not match as the size of the wireless network varies. This issue is commonly referred to as task mismatch \cite{shen2019transfer}.

Ideally, a new NN model should be trained for prediction if one or more network parameters have changed. Thus, the time-varying nature of wireless networks does impose unique challenges in using the DL-based approaches. The generality of the trained BNN is key. If the generality is sufficient, this will mean that the trained BNN can cope with a variety of dynamic situations the wireless network may face. Here we introduce two heuristic methods that can improve the generality of the DL-based approaches. The first one is the training-set augmentation method, which collects enough training samples to cover the possible changes of the network size. A large-scale model is then trained based on the augmented training set. This method works well as long as the network size is within the training set. Nonetheless, in some problems, the acquisition of a large number of samples can be too expensive. In that case, transfer learning can be used by transferring knowledge from related scenarios with additional training and labeling efforts.


\subsection*{Training-set Augmentation}
Consider the BNN for problem \textbf{P1}, as shown in Fig.~\ref{framwork_example}b, as an example which takes the channel coefficients as input and the beamforming matrix as output. With $N$ BS antennas and $K$ users, the BNN takes $2NK$ channel inputs to produce $2NK$ beamforming outputs and the factor of $2$ appears in order to handle the real and imaginary parts of the complex channel and beamforming matrices. Note that most DL tools (such as Keras and Tensorflow) only support real-valued inputs and outputs. In order to train a BNN suitable for different $N$ and $K$ values, we generate an augmented training set where the samples are diverse, i.e., the numbers of BS antennas and users can be different in different samples. To do so, the size of each sample is set to be  $2N_0K_0$ for inputs and $2N_0K_0$ for outputs, where $K_0\ge K$ and $N_0\ge N$. Thus, for each training sample with specific $K$ and $N$, there will be redundant entries at the inputs and outputs, and these entries will be filled with $0$'s. In particular, the reductant $N_0-N$ rows (or $K_0-K$ columns) of the input channel matrix, as well as the corresponding positions in the output vector, are filled with $0$'s. To achieve good generality, the training samples for different combinations of $K\le K_0$ and $N\le N_0$ should be generated with equal probabilities.


The shortcoming of the training-set augmentation method is that training a large-scale model requires a large number of training samples and takes much longer processing time. Also, the trained BNN model is still not general enough to be able to handle the cases in which $N>N_0$ or $K>K_0$.

\subsection*{Transfer Learning}

\begin{figure}
\includegraphics[width=0.47\textwidth]{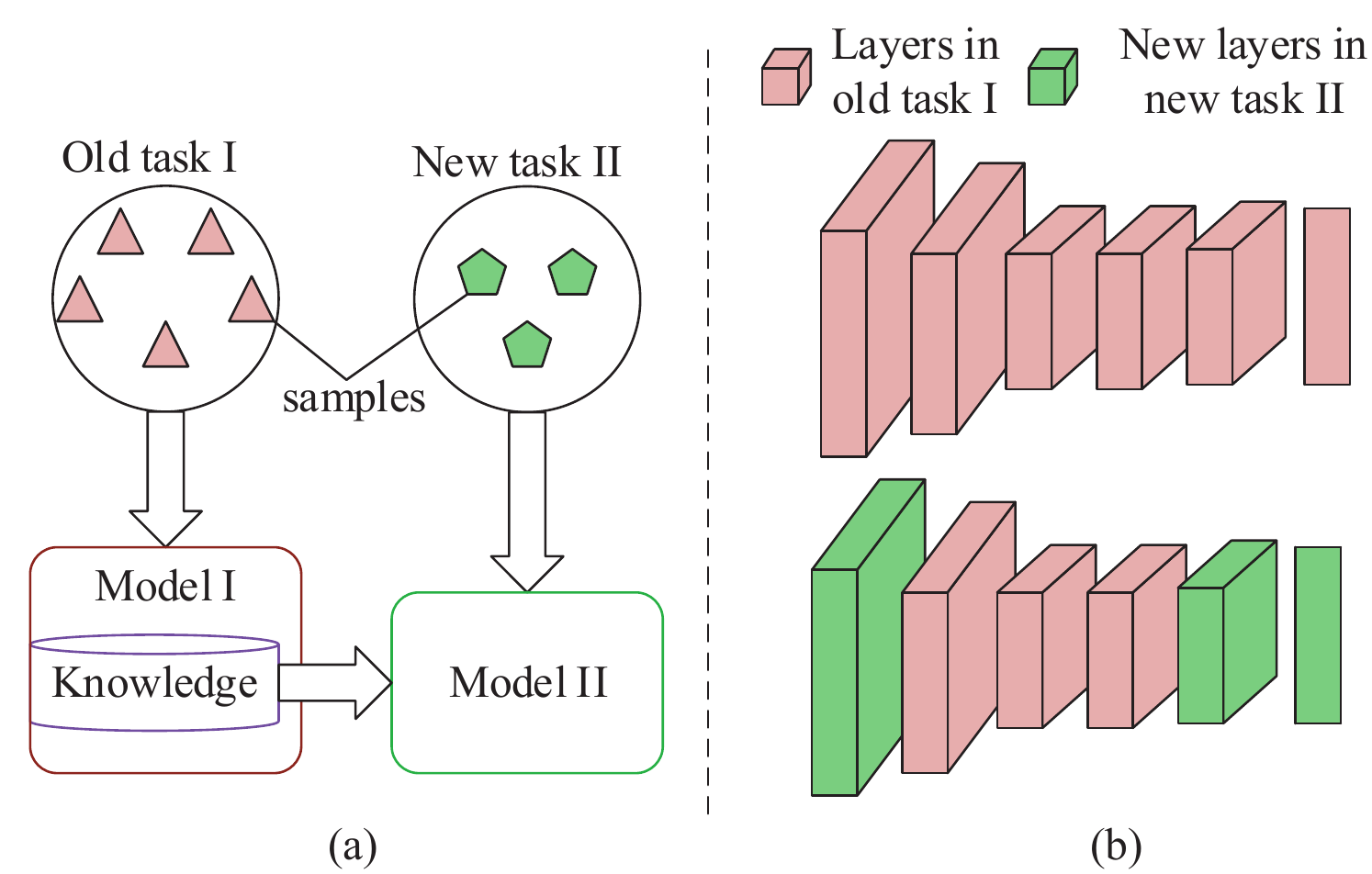}
\centering
\caption{a) Illustration of knowledge transfer and b) fine-tuning.}\label{transfer_learning}
\end{figure}

Different network settings naturally lead to different training tasks, but these tasks  share some knowledge in common about the underlying optimization problem \cite{pan2010asurvey}. As can be seen in Fig.~\ref{transfer_learning}a, knowledge learned from one training task may be transferred to another training task and can help train a new model with a few additional samples \cite{shen2019transfer}. Fine-tuning is a well-known method to implement knowledge transfer  in NNs. As shown in Fig.~\ref{transfer_learning}(b), both the input layer and output layer can be replaced by new ones suitable for the new task. Then one or more new layers are inserted into the pre-trained NN module. Fine-tuning aims to refine the pre-trained NN module with additional training samples by setting different learning rates for different layers. More specifically, we can fine-tune the newly added layers and set the learning rates of the layers from the pre-trained NN module as 0 or a very small learning rate since the initialization parameters of the pre-trained layers are expected to be a good starting point.

\section*{Performance Evaluation}

\begin{figure*}
\includegraphics[width=0.95\textwidth]{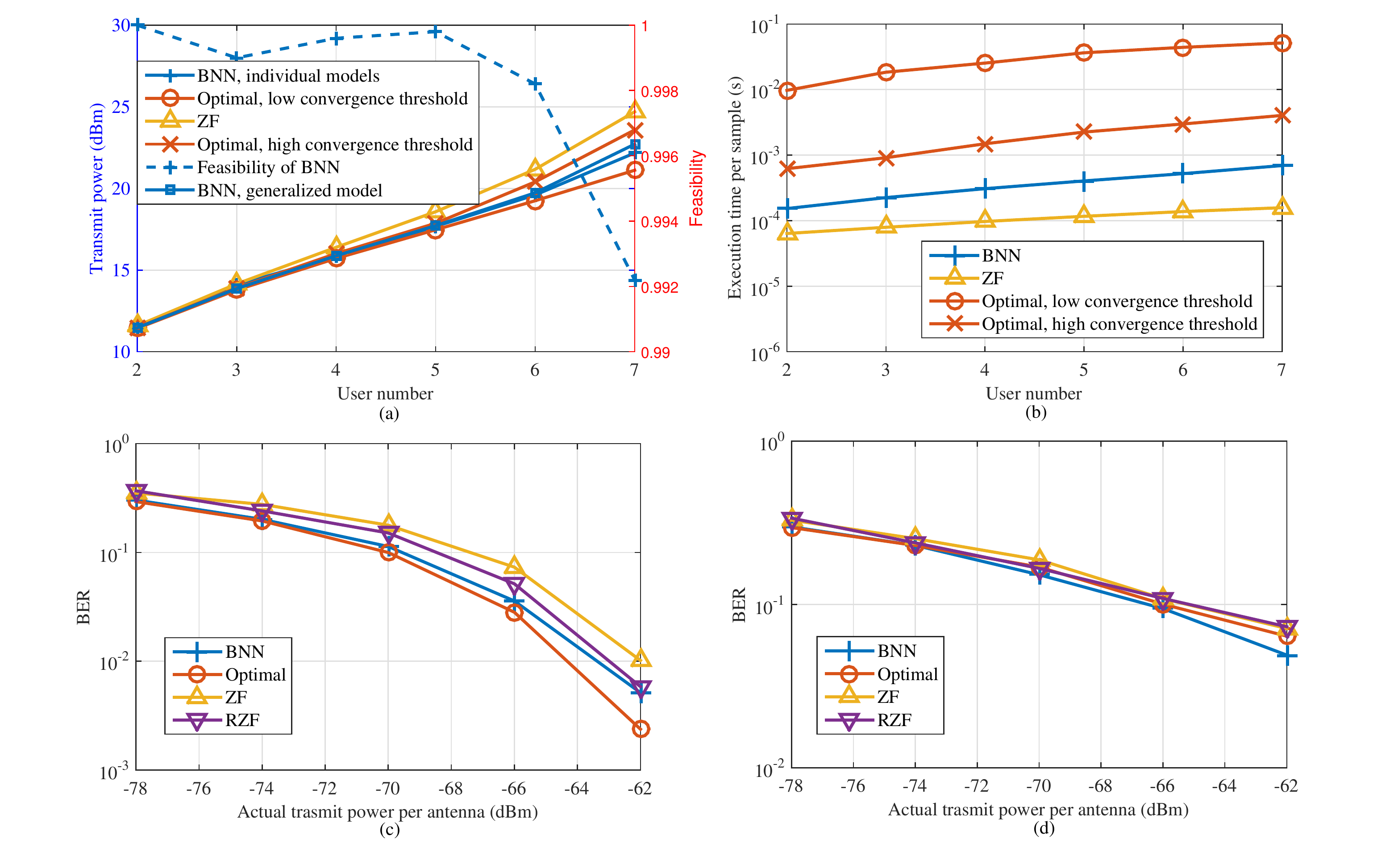}
\centering
\caption{Simulation results of the power minimization problem \cite{xia2019deep}: a) power performance and b) execution time per sample averaged over 5000 samples under QoS constraints of 5 dB and $N = 8$, and testbed  results of  the SINR balancing problem with per-antenna power constraints: c) static channel condition and d) dynamic channel condition under $K=N=4$.}\label{simulation_testbed}
\end{figure*}

To assess the performance of the DL-based BNNs, we here provide some numerical results for a downlink MISO system with $K$ users and an $N$-antenna BS using both simulations and experiments. The channel gains are generated by considering both Rayleigh fading and path loss which is modeled as $128.1 + 37.6 \log_{10}(d)~[{\rm dB}]$, with distance $d$ in km.  Also, perfect CSI is assumed available at the BS in the simulation.


We first consider the power minimization problem with QoS constraints (e.g., problem \textbf{P2} in \cite{xia2019deep}). Different from the total power constraint in problem \textbf{P1}, the QoS constraints are nonlinear and even non-convex, with which the NNs are typically not good at dealing. Moreover, due to non-zero prediction errors, the predicted results cannot always satisfy the QoS constraints. This means that there is a certain probability of infeasibility of the BNN prediction for the power minimization problem. From the results in Figs.~\ref{simulation_testbed}a and \ref{simulation_testbed}b, we compare the BNN solution to the power minimization problem  with ZF and the optimal iterative algorithm in \cite{rashid1998transmit} in terms of transmit power performance and computational delay. Note that two convergence strategies for the optimal iterative algorithm are considered: the high convergence threshold case ($10^{-2}$) which can be reached with less iterations and the low convergence threshold case ($10^{-4}$) which requires more iterations. Besides, the results marked as squares in Fig.~\ref{simulation_testbed}a correspond to the BNN solution using a generalized model with $N_0=8, K_0=7$ for the training-set augmentation method, whereas the other BNN solution is predicted based on individually trained models.

Results in Fig.~\ref{simulation_testbed}a indicate that the BNN solutions achieve better performance than the ZF beamforming and the optimal iterative algorithm for the high convergence threshold case. Results also show that the performance of the optimal iterative algorithm can be improved with more iterations, but at the cost of higher computational delay (about two orders of magnitude, compared to the BNN solutions), as seen in Fig.~\ref{simulation_testbed}b.   According to the results in Figs.~\ref{simulation_testbed}a and \ref{simulation_testbed}b, it is verified that the BNN solutions can achieve a good tradeoff between performance and complexity. Furthermore, we can find  that the feasibility of the BNN solutions for the power minimization problem can reach above  99 per cent and that the training-set augmentation method can improve the generality of the DL-based BNNs.

For the experiment results, we further take per-antenna constraints into account for the fact that in practice each transmit antenna has its own power amplifier. The SINR balancing problem with per-antenna power constraints has been investigated in many works, e.g., \cite{wang2014transmit}.
We set up the downlink MISO testbed system using software defined radio, where two NI's USRP-2950 devices are combined together to form a 4-antenna transmitter and another two USRP-2950 devices are used to emulate 4 single-antenna users.

Figs.~\ref{simulation_testbed}(c) and (d) demonstrate the bit-error rate (BER) performance against the transmit power under static and dynamic channel conditions, respectively. Under the static condition, the BNN solution outperforms the ZF beamforming and RZF beamforming. Also, it can be seen that the performance of the proposed BNN solution is inferior but close to that of the optimal solution especially in the low power regime. This is expected since the optimal solution has enough time to execute its algorithm under the static channel condition. However, this is no longer the case under the dynamic channel condition. Note that the coherence time in the environment of the experiment is 10-20 ms. In this case, although the optimal solution can achieve the best beam weights, these weights are based on the outdated CSI and therefore no longer optimal for the current CSI after long computing delay, thus leading to the performance degradation. It can be also observed in Fig.~\ref{simulation_testbed}(d) that the BER performances of the ZF solution and RZF solution with less computing time are still much worse than the proposed BNN solution.


\section*{Conclusions and Open Issues}
With the rise of DL, this article introduced the {\em model-driven} BNN solutions to beamforming optimization problems, where there is a SP module to utilize expert knowledge to empower the NN for enhanced convergence and prediction performance. We discussed the challenges of using DL for beamforming that include high dimensionality, difficulty of data acquisition for supervised learning, limited generality due to channel and network dynamics, and high prediction complexity. This article also provided methods that can improve the implementation of DL for beamforming. While there are inevitably omissions in this article, it is hoped that this article will spark interest in exploring the use of model-driven BNN for wireless communications.

It is also worth pointing out that there are some important open issues that deserve future study. For example, the first challenge is data acquisition, since generating real-world communication data is not straightforward and most existing works are based on artificial or simulated signals. It would be desirable to establish the datasets of some common problems, with which researchers can test their methods. The second challenge is how to make the DL-based BNN robust against corrupted data, which can cause inconsistency and failure of the BNN training. As the number of users increases,  it is impossible to allocate each user an orthogonal pilot and non-orthogonal pilots cause pilot contamination and CSI estimation error. Finally, it is interesting to extend the proposed method to the multi-cell distributed scenarios, which are much more complicated than the single-cell scenarios.

\bibliographystyle{IEEEtran}

\end{document}